\documentclass[a4paper, 12pt]{article}

\usepackage[sort&compress]{natbib}
\bibpunct{(}{)}{;}{a}{}{,} 

\usepackage{amsthm, amsmath, amssymb, mathrsfs, multirow, url, subfigure}
\usepackage{graphicx} 
\usepackage{ifthen} 
\usepackage{amsfonts}
\usepackage[usenames]{color}
\usepackage{fullpage}
\usepackage{tikz}
\usepackage{float}
\usepackage{appendix}
\usepackage[ruled,vlined]{algorithm2e}
\usepackage{bm}

\theoremstyle{plain}

\theoremstyle{definition}

\theoremstyle{remark} 
\newtheorem{ex}{Example}

\newtheorem{condition}{Condition}
\newtheorem{illus}{Illustration}

\newcommand{\E}{\mathsf{E}}

\newcommand{\unif}{{\sf Unif}}

\newcommand{\bet}{{\sf Beta}}

\newcommand{\RR}{\mathbb{R}}

\newcommand{\UU}{\mathbb{U}}

\newcommand{\HH}{\mathscr{H}}

\newcommand{\model}{\mathscr{P}}

\title{An Anytime Valid Test for Complete Spatial Randomness}
\author{Vaidehi Dixit\footnote{Department of Statistics, University of Missouri; {\tt vdixit@missouri.edu}, {\tt wiklec@missouri.edu}, {\tt holans@missouri.edu}}, \quad Christopher K. Wikle$^*$ \quad and \quad Scott H. Holan$^{*}$\footnote{Research and Methodology Directorate, US Census Bureau}}
\date{\today}

\begin{document}

\maketitle 

\begin{abstract}
A relevant question when analyzing spatial point patterns is that of spatial randomness. More specifically, before any model can be fit to a point pattern a first step is to test the data for departures from complete spatial randomness (CSR). Traditional techniques employ distance or quadrat counts based methods to test for CSR based on batched data. In this paper, we consider the practical scenario of testing for CSR when the data are available sequentially (i.e., online). We present a sequential testing methodology called as {\em PRe-process} that is based on e-values and is a fast, efficient and nonparametric method. Simulation experiments with the truth departing from CSR in two different scenarios show that the method is effective in capturing inhomogeneity over time. Two real data illustrations considering lung cancer cases in the Chorley-Ribble area, England from 1974 - 1983 and locations of earthquakes in the state of Oklahoma, USA from 2000 - 2011 demonstrate the utility of the PRe-process in sequential testing of CSR.

\smallskip


\end{abstract}

\section{Introduction}
\label{S:intro}
Given a spatial point pattern, an important question concerns its degree of spatial randomness. Before one can model or perform inference from such data it is imperative to know the structure of the data. For this, we are interested in testing for spatial randomness. Spatial randomness can be broadly divided in two categories, namely, global clustering and local clustering. Global clustering corresponds to the presence of clustering in the overall spatial region of interest, while local clustering is the detection/identification of clusters in subsets of the spatial domain;  this is often termed as spot detection \citep[see,][]{song2006likelihood}. Herein, we focus on global clustering. 

Methods for analyzing randomness in point patterns were initially based on first-order properties. These methods used quadrat counts that partition a spatial region $\Omega$ into $K$ sub-regions and an appropriate test statistic is defined based on $K$ quadrat counts \citep[e.g., see][]{mead1974test}. A foundation for rigorous statistical testing of CSR was given in \citet{bartlett1964spectral} based on spectral analysis of point patterns, which later led to the development of the $K$ function in \citet{ripley1976second} that was based on distances. The latter analyzed how spatial point patterns deviated from CSR when using different inter-point distances. Subsequently, spatial point patterns were analyzed for randomness in \citet{ripley1977modelling} and the power of these tests based on different distances was given in \citet{ripley1979tests}. Simultaneously, other distance-based methods were developed in \citet{diggle1976statistical}.
For a more detailed discussion of the  $K$ function, see \citet{diggle2013statistical}.

Other approaches to address randomness in spatial point patterns have been proposed. For example,  \citet{assuncao1994testing}  proposed a randomness test based on the angle between vectors joining each point to its neighbors. \citet{zimmerman1993bivariate} proposed a test based on the bivariate empirical distribution function of each location's spatial coordinates. \citet{ward2021testing} considered the problem of testing for spatial randomness on a non-Euclidean space, specifically, any three dimensional bounded convex shape of which a sphere is a particular form. Likelihood based tests for spatial clustering have been reviewed in \citet{song2006likelihood}. A related approach for testing spatial randomness is based on the notion of  scan statistics \citep{kulldorff1997spatial}. Such methods  have been proposed for a variety of point pattern structures,  with the goal being cluster detection as opposed to testing for global randomness. Discipline specific approaches to detect spatial randomness have been developed in agriculture \citep{castellazzi2007new}, botany \citep{holgate1965tests}, zoology \citep{conradt1998measuring}, epidemiology \citep{kulldorff1998statistical}, physics \citep{vavrek2008testing} and forestry \citep{corral2010permutation} among others.


Here we focus on the idea of global clustering and in particular the concept of complete spatial randomness. For a spatial point process, complete spatial randomness (CSR) is characterized by two criteria. First, the number of occurrences in any well-defined region $\Omega$ follows a Poisson distribution with mean $\lambda |\Omega|$. Given $n$ occurrences $s_1, \ldots, s_n$ in a region $\Omega$, these occurrences form an independent random sample from a uniform distribution. Thus, CSR is synonymous with a homogeneous Poisson point process. 
All existing methods for testing CSR have been developed with a goal of initial assessment/check to confirm that the available data is not CSR, before proceeding to model it by inhibition/clustering models. Hence it is assumed that the dataset is available in its entirety. This is not always the case with spatio-temporal data, where data is available as a function of time, such as disease spread, environmental occurrences, crime locations, etc. If one is interested in finding evidence for clustering or inhibition with sequential (i.e., online) data, then classical methods are inappropriate as such testing is not adaptive. For example, there could be interest in monitoring the pattern of occurrences of a disease or environmental event over time. One could repeatedly calculate the distance functions or a corresponding test statistic as new data points become available until ``evidence'' is found, but such an approach does not take ``data-peeking'' into account and can lead to erroneous conclusions. This is a common problem with any testing strategy based on {\em p-values} and recent work on {\em anytime inference} has introduced the idea of {\em e-values}, which are adaptive. 

The idea behind anytime inference is simple and easily explained in terms of betting, where the e-value is a payoff for betting against the null hypothesis. The strength of e-values is that they not only allow for data-peeking but also allow testing over multiple point patterns. For a recent review on e-values see \citet{grunwald2020safe} and \citet{shafer2021testing}. Our proposal for testing CSR is focused on the anytime approach of {\em PRe-processes} suggested by \citet{dixit2024anytime}. The strong theoretical properties and ease of computation of the PRe-process make it an attractive option. Our proposed anytime test for CSR is general in the sense that departure from the null can be tested in various directions, like a monotone decreasing pattern, a clustered process, a multivariate point pattern, or a point pattern on sphere. This is possible as the idea of the PRe-process rests on modeling the data as a mixture density and we model a flexible mixture density on the spatial point pattern.

We structure our paper as follows. Section 2 focuses on a motivating example of cancer locations where adaptive testing of CSR is beneficial followed by a formal definition of our testing problem. In Section 3, we set up our test for CSR explaining the predictive recursion algorithm and aspects of anytime inference. To illustrate the efficiency of the PRe-process we implement our methodology on three different simulation scenarios in Section 4 that exhibit the expected behavior of our test. We revisit the illustration of cancer locations in Section 5 and also showcase an application on the locations of earthquakes in the state of Oklahoma, USA. Finally Section 6 provides a concluding discussion.


\section{Motivation}
Before we set up our testing problem, we consider a spatial point pattern data where a sequential test of randomness could be of interest. With this motivation we formally state the hypotheses and our model for the alternate hypothesis of inhomogeneity.
\subsection{Illustration}
\label{ss:motivating-real}
\begin{figure}
\centering
\subfigure[]
{\includegraphics[width = 0.45\linewidth]{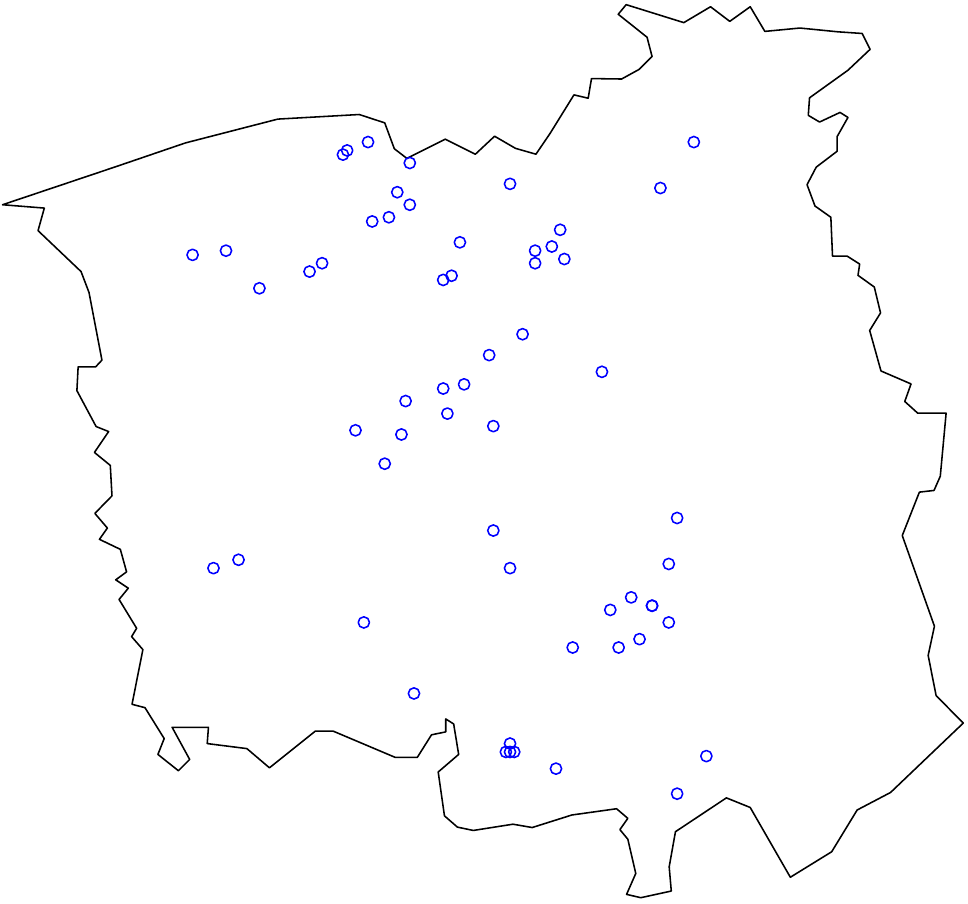}
\label{fig:chorley_larynx}}
\subfigure[]
{\includegraphics[width = 0.45\linewidth]{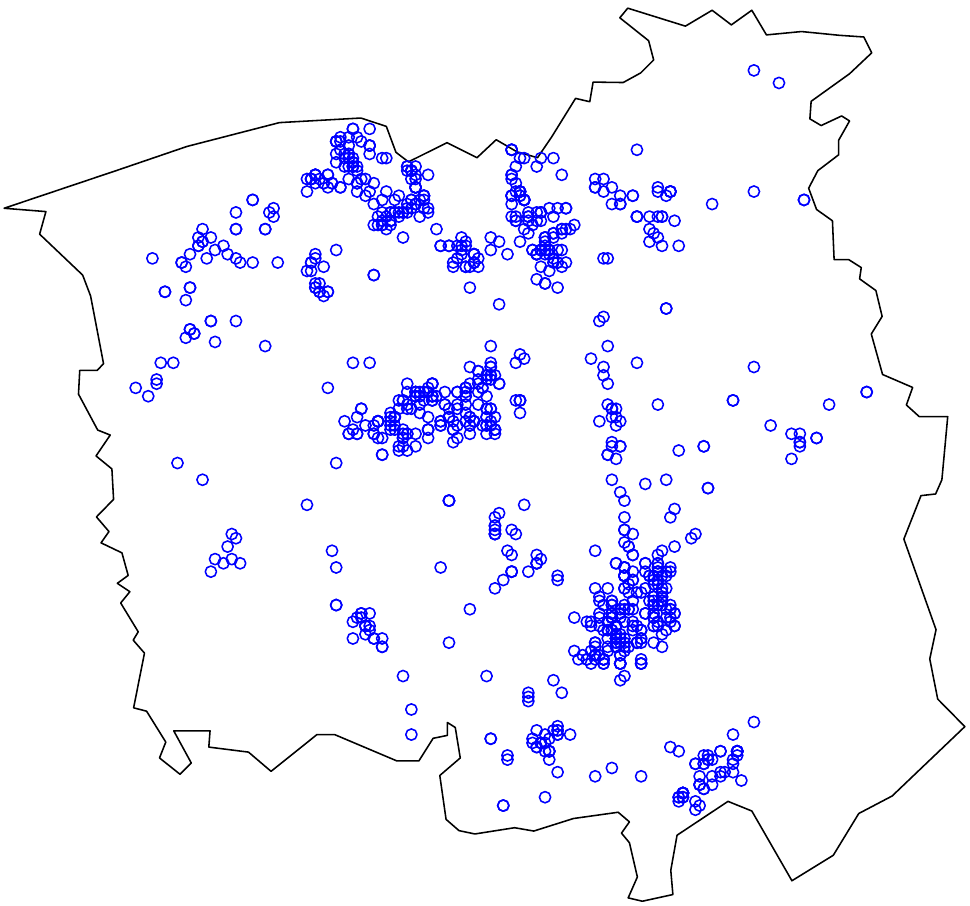}
\label{fig:chorley_lung}}
\caption{Plot of the domicile locations of (a) 58 larynx cancer cases and (b) 978 lung cancer cases in the Chorley and South Ribble Health Authority of Lancashire in England between 1974 and 1983.}
\label{fig:chorley}
\end{figure}

Consider the {\tt chorley} dataset from the {\tt spatstat} package in R \citep{baddeley2005spatstat}. The dataset consists of the precise locations of patients diagnosed with cancer of the larynx and lung in the Chorley and South Ribble Health Authority of Lancashire in England between 1974 and 1983. Figure \ref{fig:chorley} shows a plot of these point pattern. These data were first analyzed in \citet{diggle1990point} with a model for an inhomogeneous Poisson process where clustering is observed with respect to a prespecified point (in this case, a disused industrial incinerator). Data were further investigated in the spatial literature by \citet{diggle1994conditional} and \citet{baddeley2005residual}. This data set comprises 58 larynx cancer cases and 978 lung cancer cases. 

In the analyses mentioned above, the objective was to find evidence for an increase in larynx cancer cases in the vicinity of the incinerator, with the cases of lung cancer representing the susceptible population. There is a clear need to find spatial dependence in the occurrence of these cases. In our context, suppose that we observe occurrences of these cancer cases over time. Our interest might be to discover any evidence for clustering as early as possible, to quickly assess any common geographical feature contributing to the cause. To keep things simple, we focus on one type of cancer at a time. Hence, given the point pattern of, say, the lung cancer cases, we are interested in testing the hypothesis of a uniform random spread across the given geographical region versus an inhomogeneous spread. An adaptive test will update the test statistic as new cases appear and report clustering when a predetermined threshold is crossed.


\subsection{Testing problem}

The second part of the definition of {\em complete spatial randomness} (CSR) states that given $n$ occurrences $s_1, \ldots, s_n$ in a region $\Omega \subset \RR^2$, they form an independent random sample from a uniform distribution, implying homogeneity of the Poisson process. Thus, given $n$ points, we can construct our test for CSR based on the intensity function $\lambda(s)$ of a Poisson process. In this case, the  null hypothesis is defined as
\[\HH_0 : \lambda(s) \propto \unif (\Omega), \quad s \in \Omega \subset \RR^{2}.\]
To test for the inhomogeneity of the process, the alternative hypothesis should represent a robust class away from homogeneity. For example, an analyst could be interested in finding patterns of inhibition, clustering, or a non-decreasing (increasing) intensity with time. The goal is then to consider an alternative model that would be able to capture the likelihood of {\em any} of these scenarios. Let $\mathbb{P}_1$ be the alternative family.

Mixture models have been suggested in the spatial literature \citep[e.g.,][]{taddy2012mixture} as flexible models for such a scenario. To remove the influence of the range of the domain, we can assume the standardised locations to lie on $(0,1)^2$. The alternate model $\mathbb{P}_1$ can then be defined as a family $\mathscr{M}$ of bivariate beta mixture density $m(s)$, such that
\begin{equation}
\label{eq:betamix}
   m(s) = \int_\UU \bet(s \mid u) P(du), 
\end{equation}
where, $\bet(s \mid u)$ corresponds to a bivariate beta kernel density with $u=(\alpha_1, \beta_1, \alpha_2, \beta_2)$ representing the shape parameters and $P(\cdot)$ as the mixing distribution on $\UU$, where $\UU = (0,1)^2$. We can now present the test of hypothesis as
\begin{equation}
\label{eq:hyp}
    \HH_0 : \lambda(s) \propto \unif (\Omega^{*}) \quad vs. \quad \HH_1 : \lambda(s) \propto m(s), \quad s \in \Omega^{*} \subset (0,1)^{2}.
\end{equation}
Given the construction above, the question is how do we fit such a nonparametric mixture $m$ and execute the test? Nonparametric or semiparametric mixture densities have been proposed as a model for the intensity function $\lambda(s)$ in point processes \citep[see,][]{lo1989class, ishwaran2004computational, taddy2012mixture, kottas2007bayesian}. These approaches consider mixture modeling via DP mixtures or convolution of kernels with gamma processes. More recently, \citet{dixit2023prticle} suggested a quasi-Bayes approach to fitting a mixture density $m(\cdot)$ to the intensity function of a marked point pattern. The approach is the so-called predictive recursion algorithm (PR) proposed in \citet{newtonetal1998}, which is a fast, recursive algorithm for the estimation of a mixing distribution in a general mixture density $m(\cdot)$ as
\begin{equation}
\label{eq:mix}
   m(s) = \int_\UU k(s \mid u) P(du), 
\end{equation}
where $k(\cdot \mid u)$ is a known kernel density and $P$ is the unknown mixing distribution.

The particular choice of the beta kernel in \eqref{eq:betamix} is not necessary for our procedure. In fact, since the PR algorithm can accommodate any known kernel density, the fit in the alternative model can be improved for a particular pattern. For example, one could be interested in finding a monotone pattern in the occurrences and this could be nicely represented by a scaled uniform mixture \citep[see][]{williamson1955} with the kernel density as $k(x \mid u) = (1/u) 1_{[0,u]}(x)$. The improvement here is in terms of the fit of the model and in turn can also improve the efficiency of our test. We explain what we mean by efficiency in Section 3.2. If the alternative can be constructed with better knowledge of the suspected pattern then the test would be able to find evidence earlier (in terms of sample size). The beta construction we consider here is motivated by the notion that such a general mixture would be able to capture any pattern over the specified domain.
We next define the PR algorithm and explain its implementation.

\section{Test for CSR}
\subsection{Prediction Recursion}
\label{ss:pr}
Suppose we have observations $X_1, \ldots, X_n$ from a mixture density $m^{\star}$ as defined in \eqref{eq:mix}. Fitting $m^{\star}$ requires that we estimate the underlying mixing distribution $P^{\star}$. The PR algorithm is a nonparametric recursive algorithm that updates the estimate for $P^{\star}$ as each data point $X_i$ is observed. The $i^{th}$ step of this is given by
\begin{equation}
\label{eq:pr}
P_i(du) = (1-w_i) \, P_{i-1}(du) + w_i \frac{k(X_i \mid u) \, P_{i-1}(du)}{\int_\UU k(X_i \mid v) \, P_{i-1}(dv)}, \quad u \in \UU, \quad i \geq 1,
\end{equation}
where $P_0$ is the initial guess and $\{0 \leq w_i \leq 1 \mid i=1, \ldots, n\}$ is a user-defined weight sequence. In the final observation $n$, we obtain the estimate $P_n$ of the mixing distribution $P^{\star}$. Consequently, the mixture density estimate $m_n$ can be obtained by substituting $P_n$ in place of $P^{\star}$ in \eqref{eq:mix}. The weight sequence $w_i$ should satisfy $\sum_{i=1}^{n} w_i = \infty$ and $\sum_{i=1}^{n} w_i^2 < \infty$ for $P_n$ to converge. The user-defined initial guess $P_0$ may be dominated by any measure that is then carried forward to $P_n$. For almost sure consistency of $m_n$ and weak consistency of $P_n$, see \citet{martintokdar2009}, \citet{tmg2009} and \citet{dixit2023revisiting}.

For the mixture $m$ in \eqref{eq:betamix} that we consider here, the known kernel density is a joint beta density and the mixing distribution is defined over all the shape parameters of the beta kernel. The $i^{th}$ step of the algorithm requires computing a normalizing constant, i.e., $\int_\UU k(X_i \mid v) \, P_{i-1}(dv)$. This is easily achieved by the PRticle filter algorithm suggested in \citet{dixit2023prticle}, that employs an adaptive sampling approach to calculate the normalizing constant, thus enabling estimation of a multivariate mixing distribution. Given the restrictions on the weight sequence, one can choose $w_i = 1/(i+1)^\gamma$ with 
$\gamma \in [2/3, 1)$ as suggested in \citet{martintokdar2009}. A more important question in our context is restricting the support of the mixture to a geographical region. As mentioned in \eqref{eq:hyp}, we normalize the observations to lie on the unit square $(0,1)^2$, but the now-normalized geographical region may only be a small subset of this rectangle. We leverage the recursive mechanism of the PR algorithm to introduce this restriction.

Suppose $\Omega^{*} \subset (0,1)^2$ is the normalized geographical region of interest. It is not hard to generate a uniform random point pattern on $\Omega^{*}$ \citep[see {\tt rpoispp} in][]{baddeley2005spatstat}. With a reasonably large point pattern we run the PR algorithm with the beta kernel, the aforementioned weight sequence and $P_0$ as a uniform distribution over $\UU$. Given the consistency results of the PR estimate $m_n$ in \citet{dixit2023revisiting}, we expect the PR estimate $m_n$ to converge to the closest possible mixture of the form \eqref{eq:betamix} on $\Omega^{*}$. The idea here is that the PR algorithm learns through the data points about the spatial surface. With $P_n$ and $m_n$, we can now fit the mixture density in \eqref{eq:betamix} to the actual dataset of interest, but now with $P_0 = P_n$ obtained from the simulated point pattern. With this background on the PR algorithm we now present our hypothesis test based on a PRe-process.

\subsection{Anytime Inference}
\label{s:preprocess}
Here, we define our PRe-process that can be used to adaptively test the hypotheses at each data point. The PRe-process $E_n^{PR}$ from \citet{dixit2024anytime}, is given by
\begin{equation}
\label{eq:CSRe-value}
E_n^\text{\sc pr} = \frac{\hat m^\text{\sc pr}(s^n)}{\unif_{\Omega^{*}}(s^n)}, \quad n \geq 1,
\end{equation}
where $\hat m^\text{\sc pr}(s^n)$ is the PR-based marginal likelihood under the alternative and can be calculated as
\[\hat m^\text{\sc pr}(s^n) = \prod_{i=1}^{n} m_{i-1}(s^i),\]
with $m_{i-1}(s^i)$ being the PR mixture estimate of $m(s)$ based on the first $(i-1)$ observations and calculated at the $i^{th}$ data point. This expression approximates the marginal likelihood of a Bayesian Dirichlet mixture where a Dirichlet prior is assigned to the mixing distribution $P$. For additional details, see \citet{martintokdar2011}. The anytime validity and efficiency aspect of the test can now be explained.

The PRe-process in \eqref{eq:CSRe-value} is an e-process, which is based on the idea of test martingales. The foundations of this go as far back as the 18th century, where martingales were coined in the context of betting. The idea has been brought back recently \citep{grunwald2020safe, ramdas2023game, shafer2021testing, vovk2021values}. These general strategies were first constructed for a simple testing scenario based on constructing {\em e-variables}. An {\em e-variable} $E$ is simply a non-negative function of the observed data, such that $\E_{\mathbb{P}_0}(E) \leq 1$ with respect to a null probability distribution $\model_0$. The value that this e-variable takes is called the e-value, and the latter is sometimes interchangeably used to denote the former. With a traditional threshold of $\alpha$, a test that rejects $\model_0$ iff $E \geq 1/\alpha$ controls the type 1 error at level of significance $\alpha${\footnote{For some intuition, see that $1/E$ is a conservative p-value.} \citep[see Proposition 1 in][]{grunwald2020safe}. This is good for one realization of the dataset, but how does this lead to e-processes? An {\em e-process} $E_n$ for a family of distributions $\model$ is defined as a non-negative process that is bounded above by a test martingale, i.e., there exists a test martingale family $T_n^P$ for all $P \in \model$ such that,
\begin{equation}
\label{eq:e-process}
    E_n \leq T_n^P ~~\text{for all}~~ P \in \model,
\end{equation}
where $\E_{P}(T_n | F_{n-1}) = T_{n-1}$ adapted to a filtration $F_{n}$ and $T_0 = 1$.
Such an e-process $E_n$ gives an e-value $E_N$ for any stopping time $N$. Validity of an e-process is when \eqref{eq:e-process} is satisfied. This gives the control of type I error at any stopping time $N$, owing to the Ville's inequality \citep[e.g.,][Sec.~9.1]{shafer.vovk.book.2019} and a nice explanation of how it comes into play is given in Section 2.5 of \citet{ramdas2023game}. \citet{dixit2024anytime} constructed a PRe-process for a general testing problem of candidate models $\model_0$ versus 
$\model_1$. This construction is based on modeling the alternative as a mixture density via the PR algorithm. They show that the PRe-process is in fact a valid e-process (see Theorem 1 and Corollary 1 in \citet{dixit2024anytime}). Since the result is independent of the kernel density $k(\cdot \mid u)$ in the PR mixture, this results holds under the CSR testing scenario.

However, what should be the expected behavior of the e-process under the alternative? Similar to a traditional testing scenario, we expect the e-value to be large under the alternative and more importantly grow with $n$, if the alternative hypothesis is true. The efficiency of an e-process is hence explained in terms of its ``growth rate'' under the alternative \citep[see for instance,][]{ramdas2023game}. \citet{dixit2024anytime} show that the general PRe-process is asymptotically growth optimal under the alternative hypothesis, under some suitable regularity conditions. In our spatial context, we consider asymptotics to be of the infill type \citep[e.g., see][]{stein1999interpolation}. We establish the growth rate optimality for the specific PRe-process below, for these conditions. The conditions under which the growth rate optimality is obtained are split into two parts. One concerns the consistency of the PR estimate $m_n$ and the other the features of the null hypothesis. The latter assumes a bound on the ratio of the true density and the maximum likelihood estimate under the null, i.e.,
\begin{equation}
\label{eq:mle.limit}
\liminf_{n \to \infty} n^{-1} \log\{ m^\star(X^n) / \hat m_0(X^n)\} \geq \kappa^\star(\model_0), \quad \text{with $P^\star$-probability 1}. 
\end{equation}

Let us consider the conditions for the consistency of the PR estimate $m_n$ when defined for the general mixture model in \eqref{eq:mix}.
Suppose that $m^{\star} (s)$ denotes the true probability density, which need not be a mixture. Also let $m_{p} (s)$ be the `best projection' of the true $m^{\star}$ in our mixture family $\mathscr{M}$, where we define the best projection in terms of the Kulback-Leibler divergence, i.e.,
\begin{equation}
\label{a:eq:inf}
K(m^\star, m_p) = \inf_{m \in \mathscr{M}} K(m^\star, m). 
\end{equation}
The conditions required for the consistency of $m_n$ are given below. 

\begin{condition}
\label{cond:support}
$\UU$ is compact and $u \mapsto k(x \mid u)$ is continuous for almost all $x$. 
\end{condition}

\begin{condition} 
\label{cond:weights}
The PR weight sequence $w_i$ satisfies 
\begin{equation}
\label{eq:weights}
\sum_{i=1}^\infty w_i = \infty \quad \text{and} \quad \sum_{i=1}^\infty w_i^2 < \infty. 
\end{equation}
\end{condition}

\begin{condition}
The kernel density $k(x \mid u)$ satisfies
    \label{cond:integrable}
    \begin{equation}
    \sup_{u_1, u_2 \in \UU} \int \Bigl \{\frac{k(x \mid u_1)}{k(x \mid u_2)} \Bigr \}^{2} \, m^\star(x) \, dx < \infty
    \end{equation}
\end{condition}

\begin{condition}
\label{cond:condition4}
The Kullback--Leibler projection $m_p$ in \eqref{a:eq:inf} satisfies
\begin{equation}
\label{eq:condition4}
\int \Bigl\{ \log \frac{m^\star(x)}{m_{p}(x)} \Bigr\}^2 \, m^\star(x) \, dx < \infty. 
\end{equation}    
\end{condition}

We explain that these are satisfied for our particular mixture construction in \eqref{eq:betamix}. Condition 1 can be satisfied by choosing a sufficiently large but compact support $\UU$ for the beta parameters $\alpha_1, \alpha_2, \beta_1, \beta_2$. Condition 2 is satisfied by the choice of our weight sequence $w_i = 1/(i+1)^\gamma$ with $\gamma \in [2/3, 1) ~~\forall ~i$. Condition 3 is satisfied for the beta kernel $k(s \mid u) = \bet (s \mid u)$ as the ratio $[{k(x \mid u_1)}/{k(x \mid u_2)}]$ is bounded for all $u_1, u_2 \in \UU$, where $\UU$ is compact. Finally for condition 4 to be satisfied, we need $\log m^{\star} (x) / m_{P}(x)$ to be bounded, which means that the mixture family we choose is adequate in covering all the support for $m^{\star}$. Since $m^\star$ will be defined on a compact domain (geographical region), the mixture family \eqref{eq:betamix} we choose can adequately cover all support. Thus, we have consistency of the PR estimate $m_n$. We present the steps for implementing this test for CSR in Algorithm~\ref{algo:CSR} given in the Appendix.

\section{Simulation study}
To illustrate the performance of the PRe-process testing strategy we conduct three simulation studies. We first show that the PRe-process in \eqref{eq:CSRe-value} will in fact decrease towards zero if the true point process is indeed a homogeneous Poisson point process. In the other two simulation studies we choose the truth to have two different kinds of inhomogeneities. The first inhomogeneous process we consider is the Matern process, which is a simple Neyman-Scott process, and the second as a point process with a monotonically decreasing intensity function. Without loss of generality, all point processes are standardized to lie on a $(0,1)^{2}$ rectangle. For all experiments, we generate 100 datasets from the truth. 

To exhibit the anytime aspect of the methodology, we assume that the data are received in batches over time and the PRe-value is calculated at regular intervals.   For calculation of the numerator in the PRe-process, we use the PRticle filter approach given in \citet{dixit2023prticle}. This is displayed in Algorithm \ref{algo:prticle} in the Appendix. Under this, $U$ is initialized with particles of size $t=10000$ such that each shape parameter in $u=(\alpha_1, \beta_1, \alpha_2, \beta_2)$ is generated from a $\unif(0.2, 10)$. The likelihood in the numerator of \eqref{eq:CSRe-value} is obtained as a result of the recursive calculations of PR. Given its anytime property, recall that each calculation of the PRe-value takes into account the previous calculation and simply updates that based on the new data available.  

\begin{ex}
First, consider the true point process as completely spatially random. We generate 100 datasets from a homogeneous Poisson process on a $(0,10)^{2}$ window with intensity $\lambda = 10$ and standardize the locations to lie on $(0,1)^{2}$. Such a dataset contains approximately $1000$ datapoints. A plot of one such dataset is given in Figure \ref{fig:dataset_homogeneous}. For each dataset, the PRe-value is calculated at steps of $100$ with the view that the points are available in batches. A plot of $\log E_n^\text{\sc pr}$ vs $n$ is given in Figure \ref{fig:evalue_homogeneous}. Clearly all $\log E_n^\text{\sc pr}$ values are below the rejection threshold of $\log{1/\alpha}$ with $\alpha = 0.05$ and the PRe-value decreases towards zero, indicating no evidence for the alternative.

\begin{figure}
\centering
\subfigure[]
{\includegraphics[width = 0.45\linewidth]{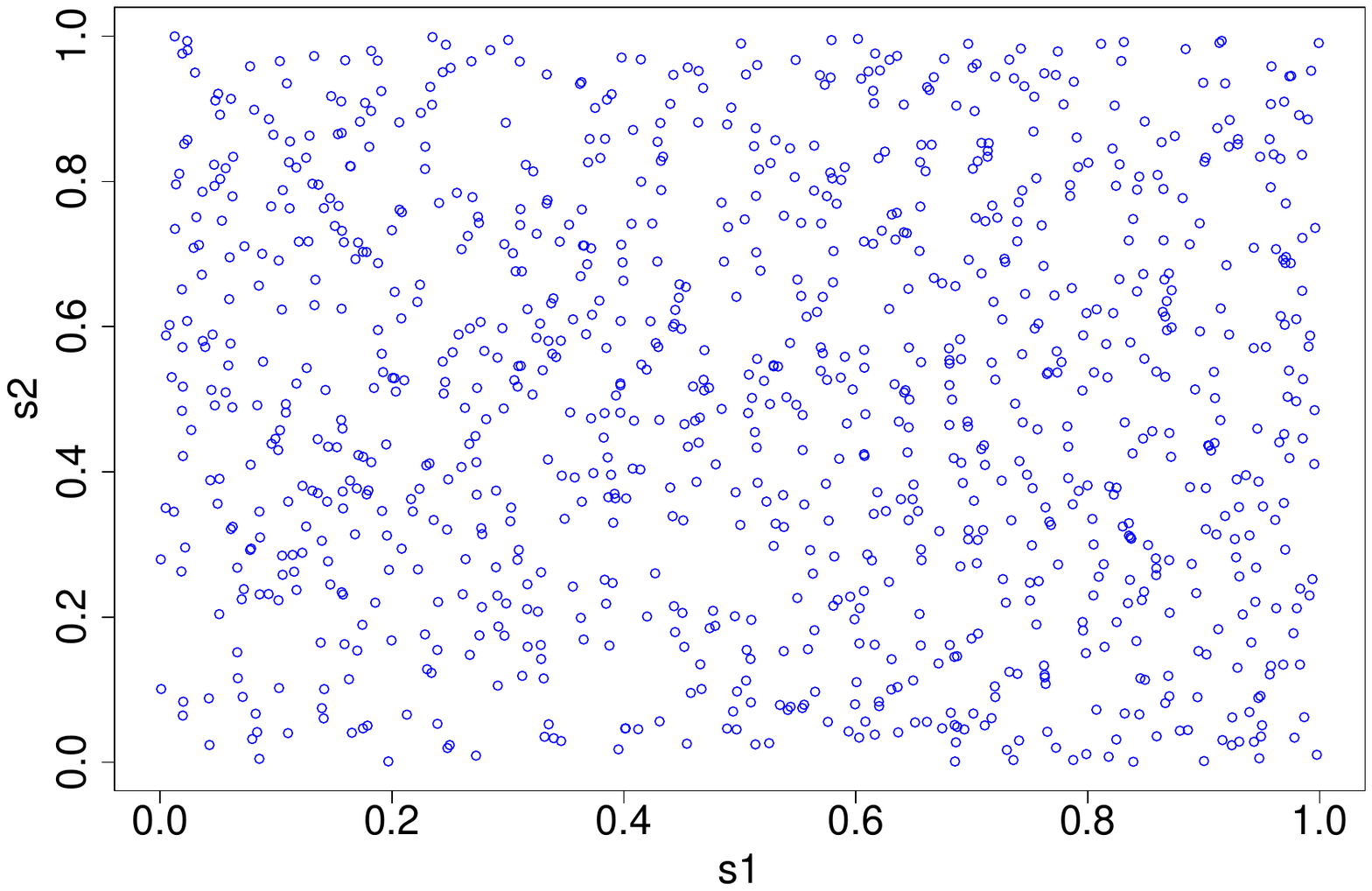}
\label{fig:dataset_homogeneous}}
\subfigure[]
{\includegraphics[width = 0.45\linewidth]{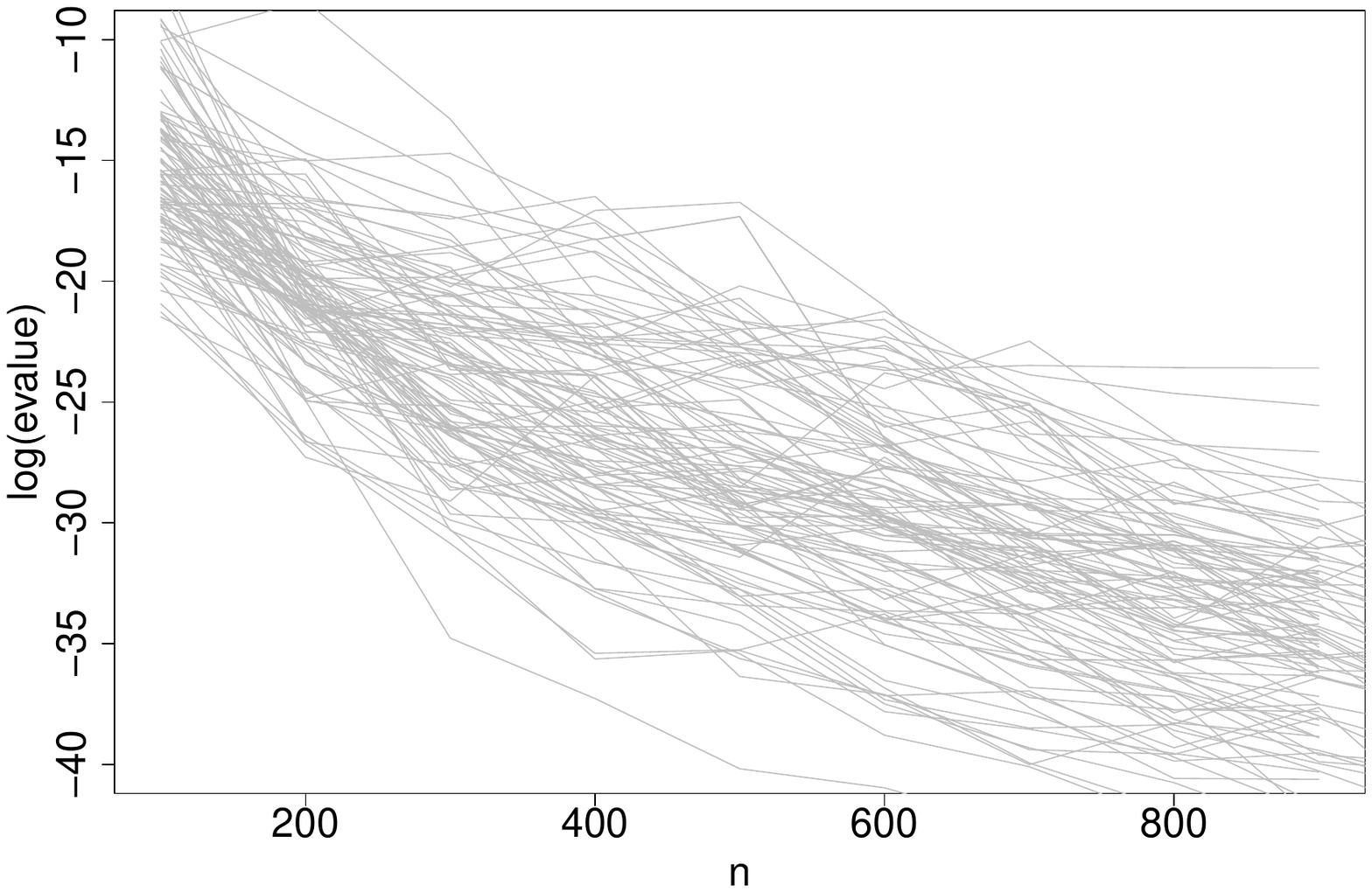}
\label{fig:evalue_homogeneous}}
\caption{(a) Single dataset from a homogeneous Poisson process with $\lambda = 10$ (b) Plot of $\log E_n^\text{\sc pr}$ vs $n$ for 100 datasets generated from this process.}
\label{fig:homogeneous}
\end{figure}
\end{ex}

\begin{ex}
In the second experiment, we generate observations from a Matern process with $\kappa = 50$, scale = 0.1 and $\mu = 20$ in the window $(0,1)^2$. The generation happens in two steps -- first the parent points are generated from a homogeneous Poisson point process on $(0,1)^2$ with intensity $\kappa$, then with these points as centers of discs with radius equal to scale, each parent point is replaced by a homogeneous Poisson process with intensity $\mu$. The resulting pattern for one dataset is given in Figure~\ref{fig:matern}. Such a dataset has approximately 1000 observations. As per the simulation characteristics described previously, we generate 100 such datasets and calculate the PRe-value at steps of $100$. We show a plot of $\log E_n^\text{\sc pr}$ vs $n$ for all $100$ datasets in Figure~\ref{fig:maternevalue}. In terms of frequentist rejection, the rejection criteria here is that if $\log E_n^\text{\sc pr} > \log (1/\alpha)$, we reject the null hypothesis. This can be interpreted as evidence found against the null and happens very early on ($n<100$) in the data sequence for all datasets. This aligns with the results obtained from running a {\tt Ktest} (based on the $K$-function) on the same data. In the anytime inference framework, it is possible to achieve perfect power as the sample size increases. This is due to the growth property of the e-processes that allows for accumulated evidence.
To further emphasize the anytime aspect of the testing strategy, we conduct a side experiment. We generate the first $300$ observations from the aforementioned Matern process and the next $800$ observations as uniformly distributed over the $(0,1)^2$ domain. We calculate the PRe-value as before at steps of $100$. Under this, one can see that the trend of the PRe-value changes around the $n=300$ mark, indicating change in evidence as shown in Figure \ref{fig:matern_uniform}. This is an important result and only possible in an anytime testing scenario. In other words, the PRe-process is able to change course when the pattern strays away from clustering. This warrants further investigation and in fact \citet{ramdasdetectors} construct an anytime e-detector based on the e-process in \cite{wasserman2020universal} that is able to detect a change-point in data. Since this is out of the scope of the current topic we refer to this later in the discussion section.

\begin{figure}
\centering
\subfigure[]
{\includegraphics[width = 0.45\linewidth]{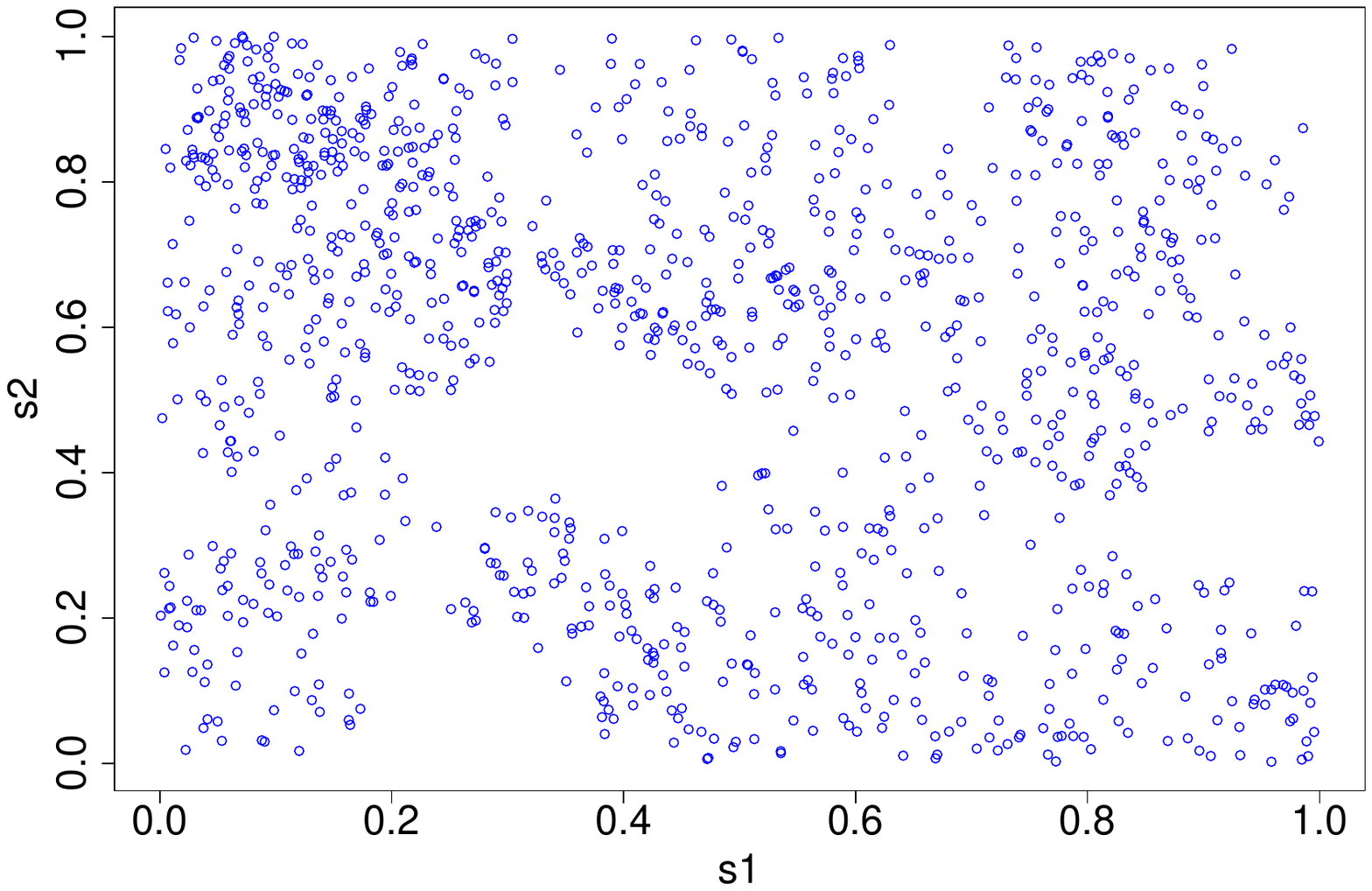}
\label{fig:matern}}
\subfigure[]
{\includegraphics[width = 0.45\linewidth]{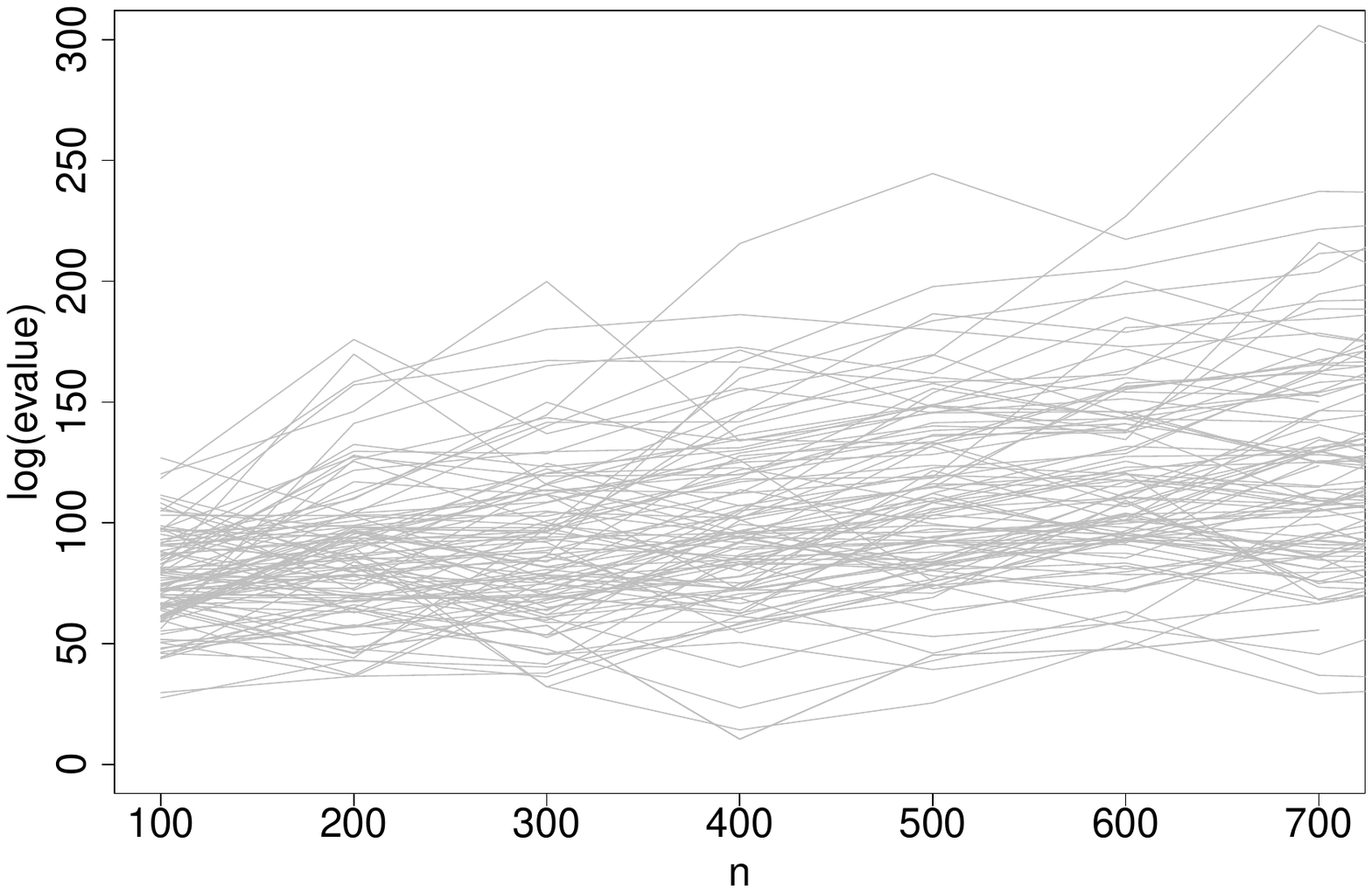}
\label{fig:maternevalue}}
\caption{(a) Single dataset from $Matern(\kappa = 50, scale = 0.1, \mu = 20)$ (b) Plot of $\log E_n^\text{\sc pr}$ vs $n$ for 100 datasets generated from a $Matern(\kappa = 50, scale = 0.1, \mu = 20)$ process.}
\end{figure}

\begin{figure}
    \centering
    \includegraphics[width = 0.75\linewidth]{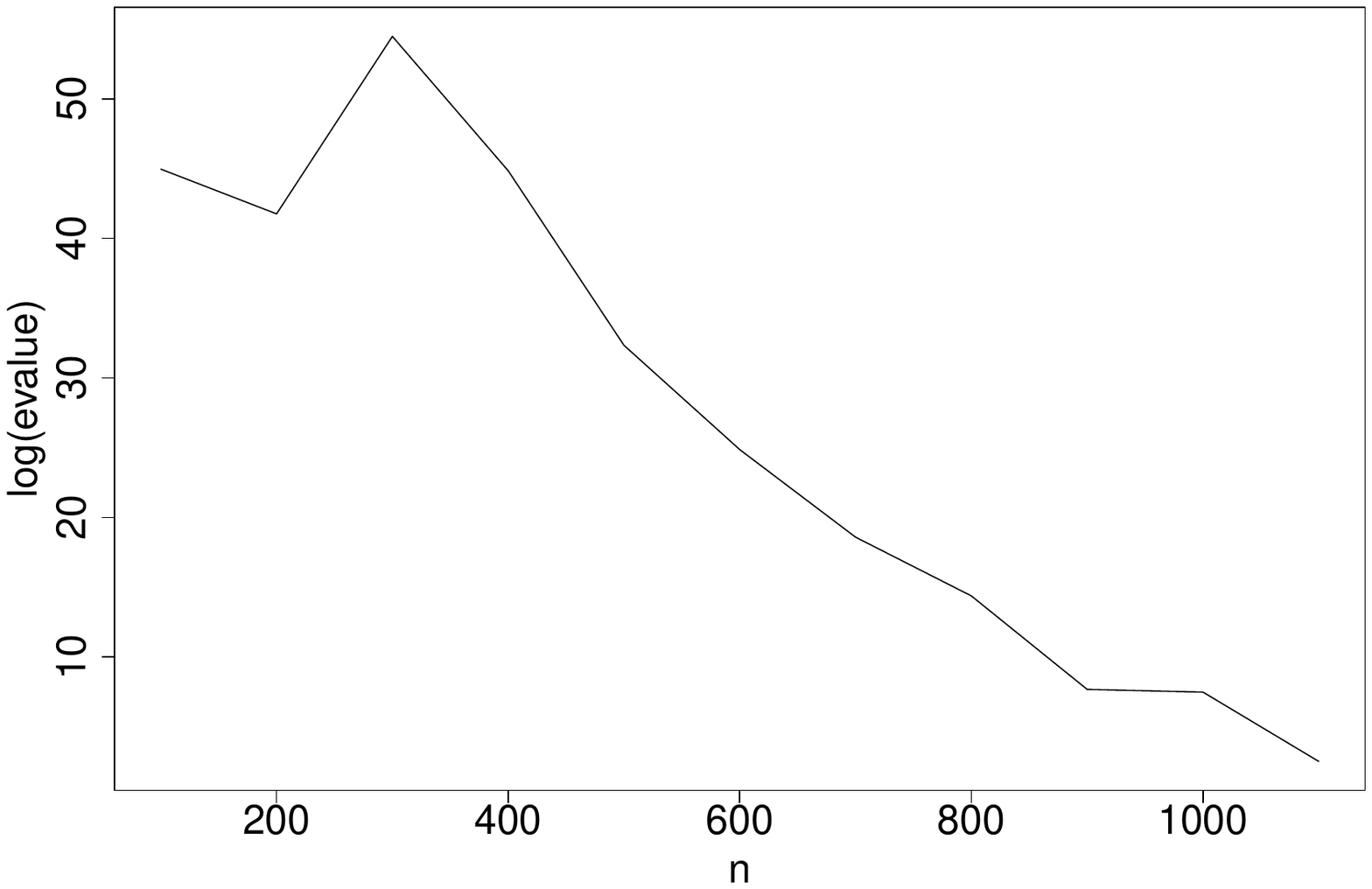}
    \caption{Plot of $\log E_n^\text{\sc pr}$ vs $n$ for a single dataset, where the initial $300$ observations are from a Matern process while the latter $800$ are uniformly distributed.}
    \label{fig:matern_uniform}
\end{figure}
\end{ex}

\begin{ex}
For the third simulation experiment we consider an inhomogeneous Poisson process that has the intensity function $\lambda(s_1, s_2)$ as a truncated joint exponential distribution, i.e.,
\begin{equation}
\label{eq:exp_intensity}
    \lambda(s_1, s_2) = \lambda_0 \frac{\gamma_1 \gamma_2}{(1 - \exp{(-\gamma_1)})(1 - \exp{(-\gamma_2}))} \exp{(-\gamma_1 s_1)} \exp{(-\gamma_2 s_2)}, 
    \quad 0 < s_1,s_2 < 1
\end{equation}
with $\lambda_0 = 1000$ and consider two cases for $(\gamma_1, \gamma_2)$, i.e.,$(2,4)$ and $(10,10)$. The exponential function helps to define an intensity that is monotone in one or both dimensions. The two cases help to exhibit the change in the PRe-process when the monotonicity of the intensity increases. The plots of the intensity function for one data set from each of the two cases are given in Figure \ref{fig:monotone_dataset}.

\begin{figure}
\centering
\subfigure[]
{\includegraphics[width = 0.45\linewidth]{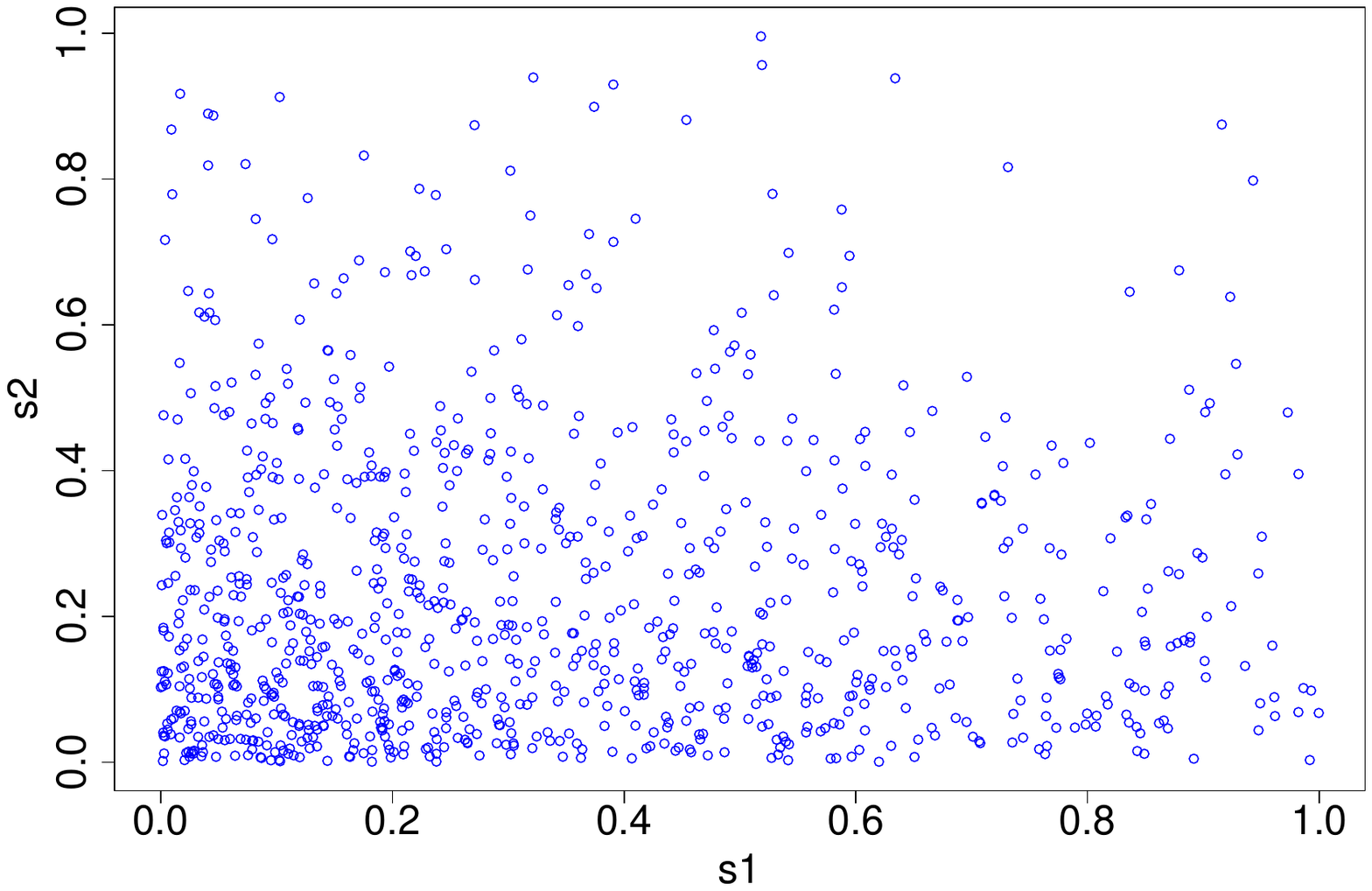}
\label{fig:monotone1}}
\subfigure[]
{\includegraphics[width = 0.45\linewidth]{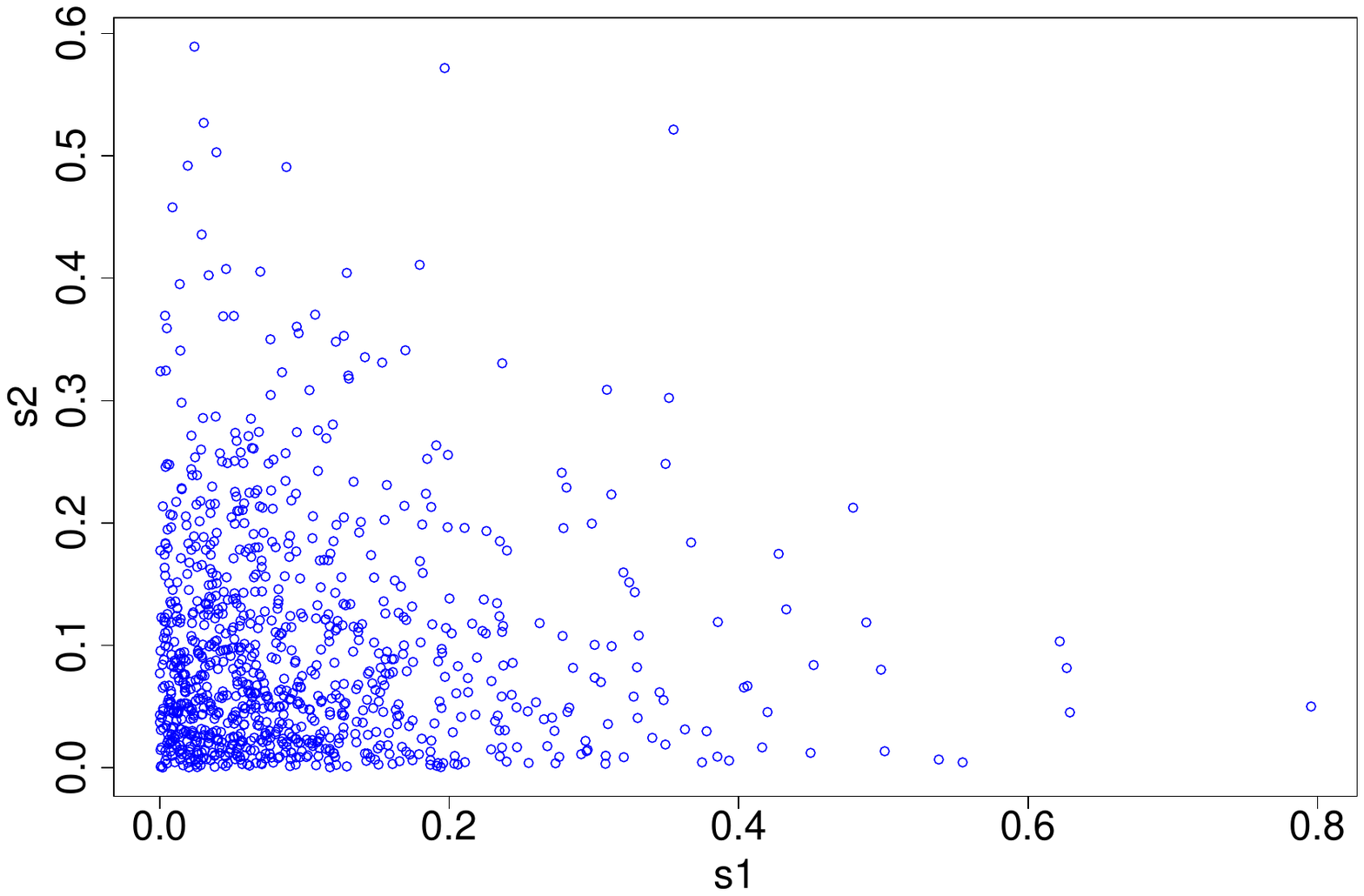}
\label{fig:monotone2}}
\caption{(a) Single dataset from a joint truncated exponential distribution \eqref{eq:exp_intensity} with (a)$(\gamma_1, \gamma_2) = (2, 4)$ on $(0,1)^2$ and  (b) $(\gamma_1, \gamma_2) = (10, 10)$ on $(0,1)^2$.}
\label{fig:monotone_dataset}
\end{figure}

We repeat the construction of the PRe-process in the same way as before and a plot of $\log E_n^\text{\sc pr}$ vs $n$ is given in Figure \ref{fig:monotone_evalues} for each of the two cases. A frequentist interpretation would reject the null hypothesis of CSR for all $n$ as all $\log E_n^\text{\sc pr}$ values are above the threshold of $\log (1/\alpha)$ for $\alpha = 0.05$. Also, as we go from Case 1 to Case 2, we see that the slope of the growth is higher for Case 2 which is consistent with the fact that Case 2 has {\em stronger} inhomogeneity than Case 1.  Further, since the true intensity function is known in this case, we calculate the theoretical growth rate and overlay this on both plots. The simulated PRe-processes adhere to this expected growth.
\begin{figure}
\centering
\subfigure[]
{\includegraphics[width = 0.45\linewidth]{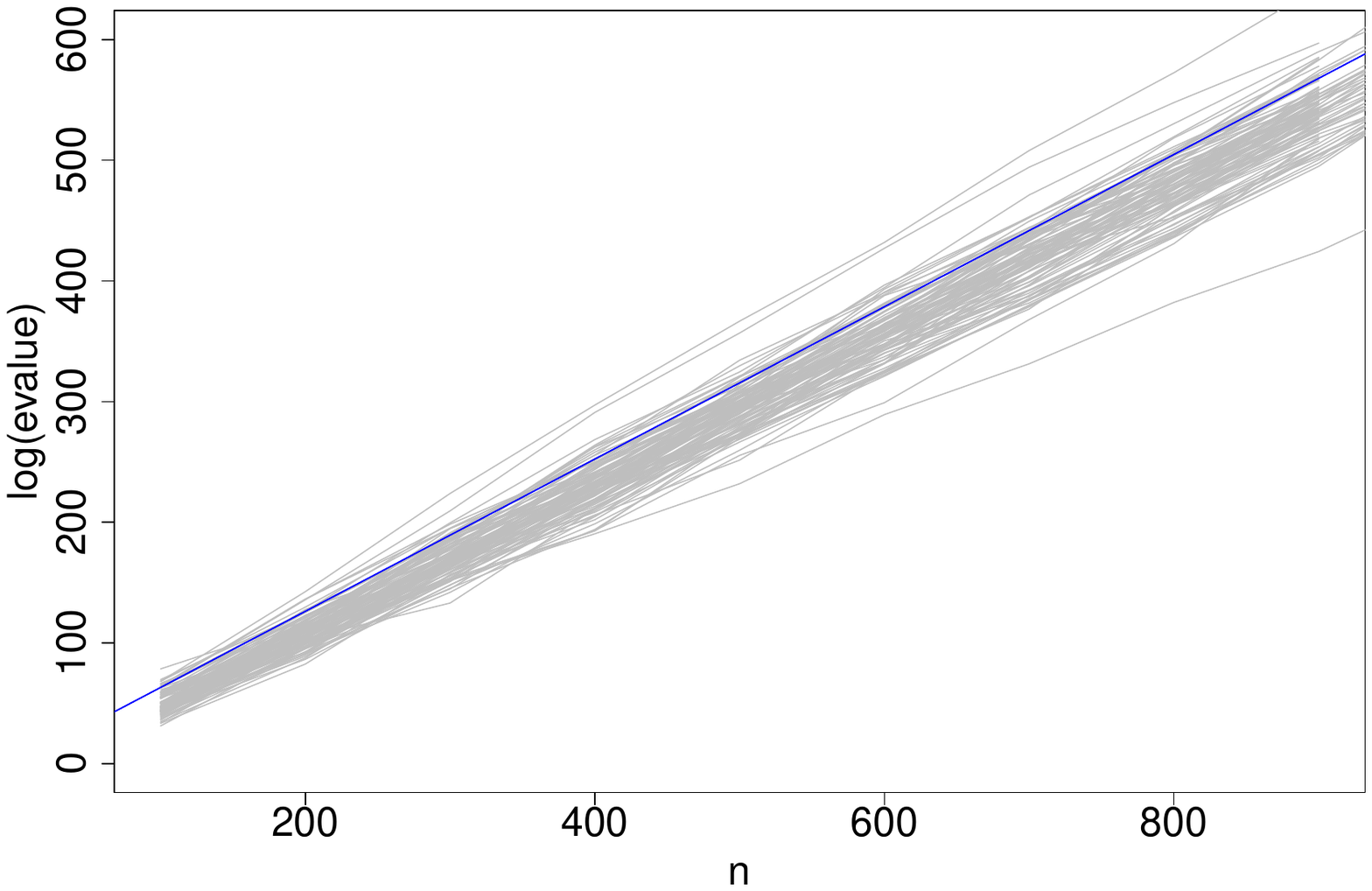}
}
\subfigure[]
{\includegraphics[width = 0.45\linewidth]{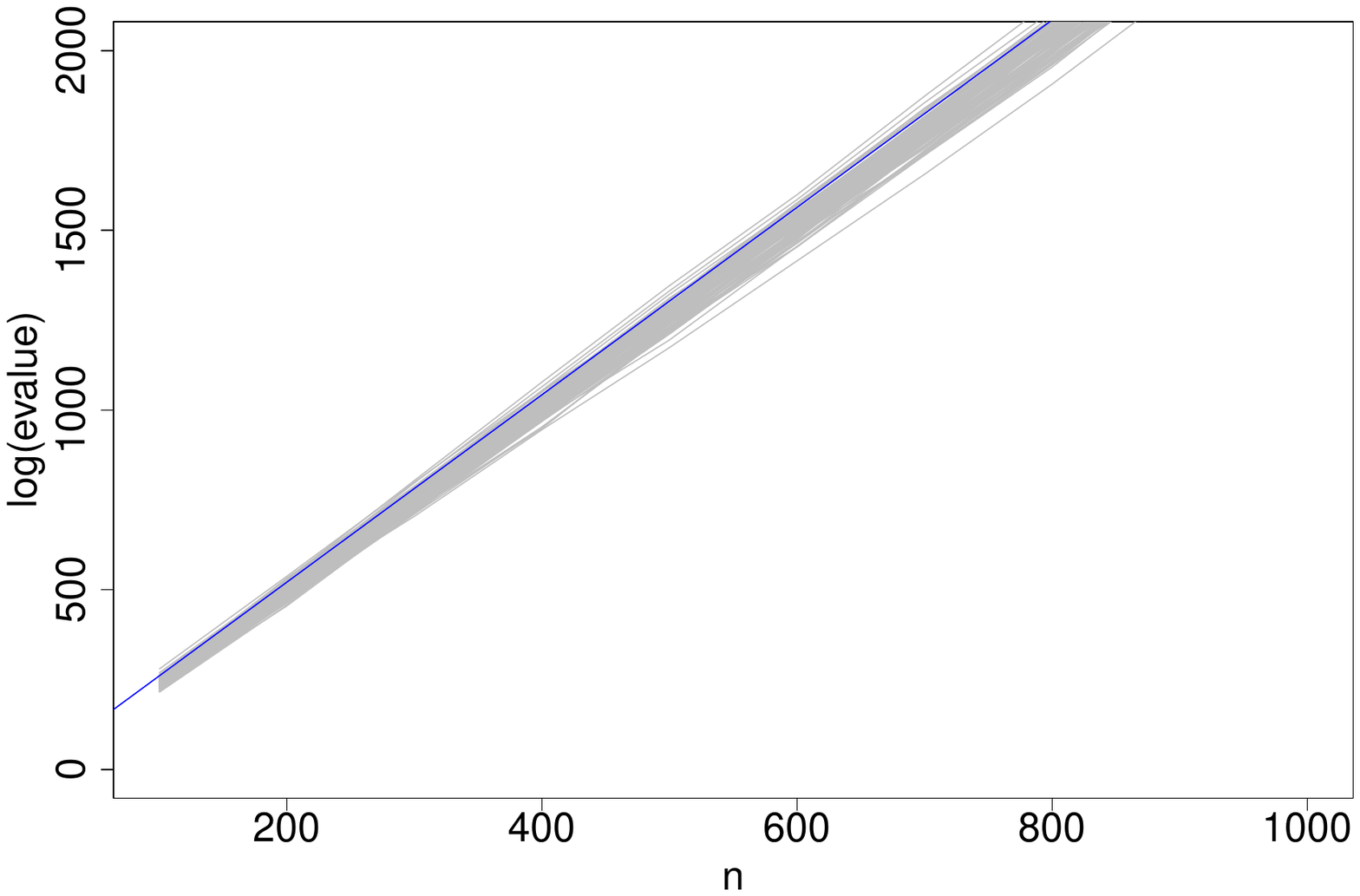}
}
\caption{Plot of $\log E_n^\text{\sc pr}$ vs $n$ for $100$ datasets generated from a a joint truncated exponential distribution for two cases as in Figure \ref{fig:monotone_dataset}. Both plots are overlaid with the respective theoretical growth rate (blue).}
\label{fig:monotone_evalues}
\end{figure}
\end{ex}

\section{Data Illustration}
Given the theoretical and empirical evidence we have, we now conduct an investigation of the spatial pattern of two real datasets. One is the motivating dataset of lung cancer cases in the Chorley-Ribble area given in Section \ref{ss:motivating-real} and the other is the pattern of earthquakes in Oklahoma. To test for spatial randomness, we standardize the observations to lie on a rectangular window of $(0,1)^2$. Since the region of interest is only a subset of the rectangular window, we need to ensure this restricted support in our calculations. For the mixture density, we first generate a homogeneous pattern over the standardized region and run a trial of the PR algorithm with $U$ initialized as particles of size $t=10000$ such that each shape parameter in $u=(\alpha_1, \beta_1, \alpha_2, \beta_2)$ is generated from a $\unif(0.2, 10)$. This gives us a first fit of a mixture density, as described at the end of Section \ref{ss:pr}. With $P_n$ and $m_n$, we construct the numerator of the PRe-process. For the denominator of the PRe-process we need to fit a uniform distribution on a subset of the $(0,1)^2$ space. Hence, the expression of $\unif_{\Omega^{*}}(s^n)$ can be calculated as $1/(\text{area of subset})$. 

\begin{illus}
Let us first consider the Chorley-Ribble dataset as described in Section \ref{ss:motivating-real}. The goal here is to find evidence for clustering in the point pattern of lung cancer cases over time. One caveat here is that the dataset does not give the time stamp of when the cases were observed, but is available as a batch. This gives us an opportunity to use the strength of the PRe-process, over different orderings of the data. For this exercise, we generate $100$ permutations of the lung cancer cases indicating different ordering of occurrence and run the PRe-process setup over each of these permutations. The idea is to see how the PRe-process would change as the ordering of the datapoints change. Figure \ref{fig:chorley.lung.evalue} gives a plot of 
$\log E_n^\text{\sc pr}$ vs case number for these 200 permutations. We can see that the varying pattern results into a varying PRe-process, however with the same underlying trend. For another interpretation, we calculate an empirical rejection probability with $\alpha = 0.05$, i.e. the proportion of permutations for which the threshold of $\log (1/0.05)$ is crossed. A plot of this is given in Figure \ref{fig:chorley.rej.prob}. The message here is clear. When the point pattern does not cluster with initial observations, evidence for inhomogeneity is not found early, but in some permutations it is achieved as early as at $n=200$. This showcases the adaptive testability of the PRe-process. 

\begin{figure}
\centering
\subfigure[]
{\includegraphics[width = 0.45\linewidth]{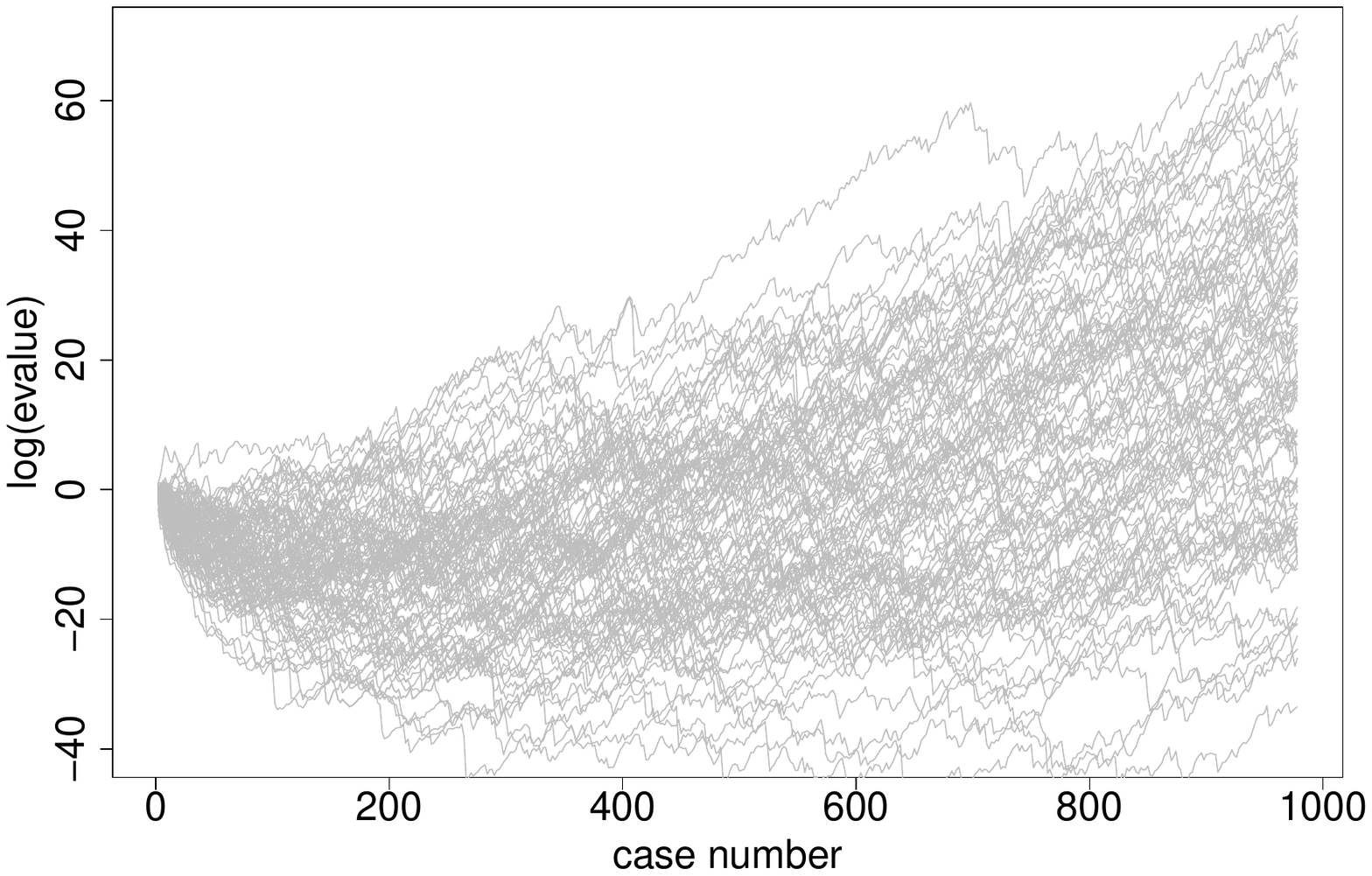}
\label{fig:chorley.lung.evalue}}
\subfigure[]
{\includegraphics[width = 0.45\linewidth]{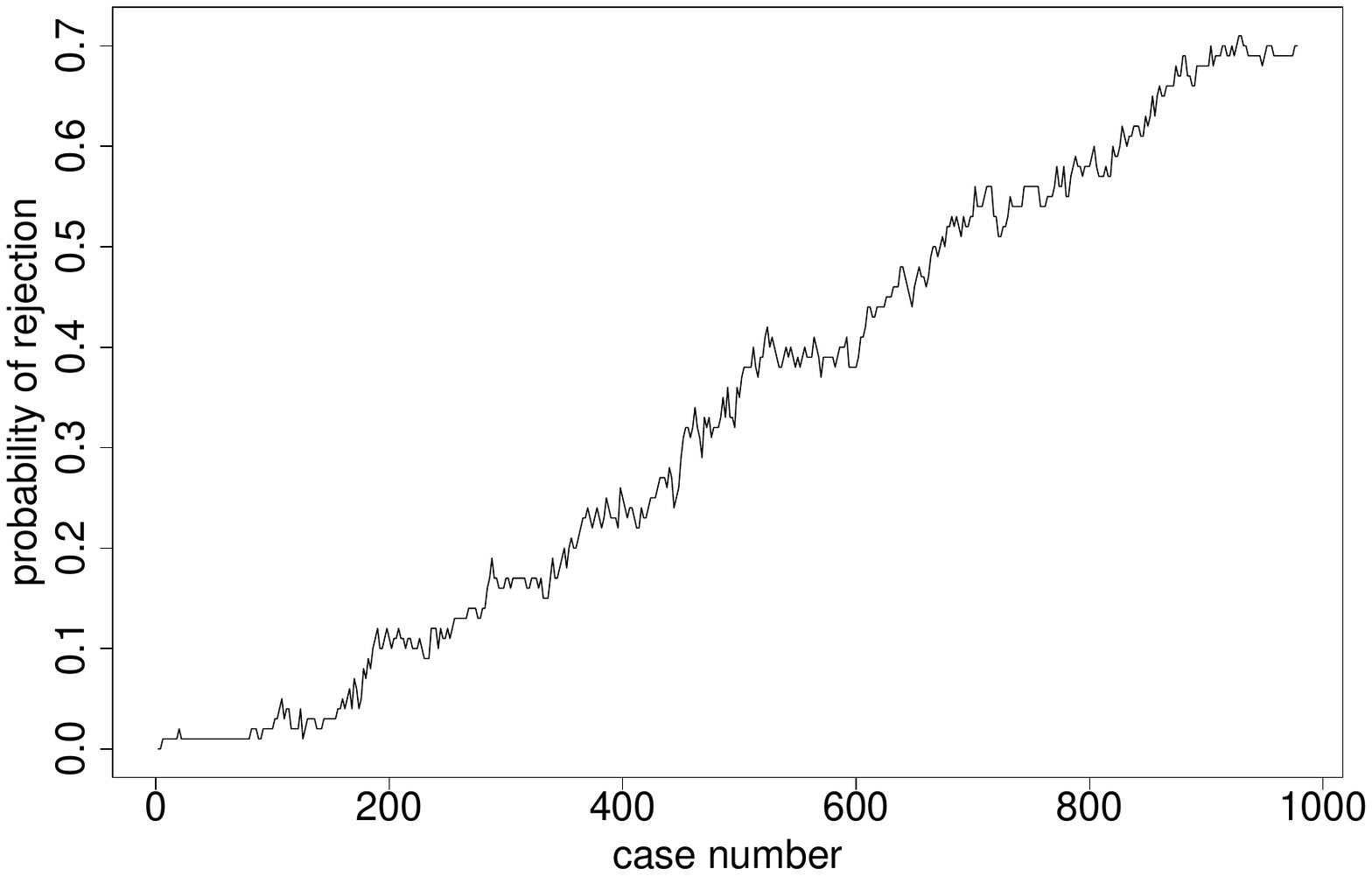}
\label{fig:chorley.rej.prob}}
\caption{(a) Plot of $\log E_n^\text{\sc pr}$ vs $\text{case number}$ for $100$ permutations of the domicile locations of $978$ lung cancer cases in the Chorley-Ribble area between 1974-1983. (b) Plot of proportion of permutations for which the threshold of $\log (1/0.05)$ is crossed at case numbers $n=2,4,\ldots, 978$.}
\end{figure}
\end{illus}

\begin{illus}
The state of Oklahoma has had the most number of induced earthquakes in the United States \footnote{https://www.usgs.gov/news/national-news-release/century-induced-earthquakes-oklahoma}. These are suspected to be a result of fracking (extraction of oil and gas from rock formations) or wastewater disposal. The year-wise locations of earthquakes in Oklahoma accompanied by their magnitude are provided by the Oklahoma Geological Survey
\footnote{\tt https://www.ou.edu/ogs/research/earthquakes/catalogs}. With this motivation, we investigate the locations of earthquakes in Oklahoma from the year 2000 to 2011 to find evidence of clustering over time. There were a total of 2871 earthquakes in our dataset as shown in Figure \ref{fig:earthquakes}. We consider all earthquakes with a magnitude over 3.0, resulting in a total of $141$ earthquakes (after removing an outlier in the pan-handle). The frequency of earthquakes increases dramatically from 2009, with the cumulative earthquakes in $2008$ being $13$ and that in $2011$ being $141$. Here the PRe-value is updated as new locations of earthquakes are observed, and we display the trend of $\log E_n^\text{\sc pr}$ in two ways. First, we plot $\log E_n^\text{\sc pr}$ vs case number in Figure \ref{fig:earthquake_3_all}, and second we plot $\log E_n^\text{\sc pr}$ vs year in Figure \ref{fig:earthquake_3_years}. By frequentist rejection, the null hypothesis of CSR is rejected from the year 2002 onwards for $\alpha = 0.05$. The growth in $\log E_n^\text{\sc pr}$ is slower at first but increases sharply from case number $40$. In the year-wise plot (Figure \ref{fig:earthquake_3_years}) a sharp increase can be seen at year 2009 consistent with the USGS findings \footnote{https://www.usgs.gov/faqs/oklahoma-has-had-a-surge-earthquakes-2009-are-they-due-fracking}, and is indicative of larger number of earthquakes contributing to the clustering pattern from that year.

\begin{figure}
\centering
\subfigure[]
{\includegraphics[width = 0.80\linewidth]{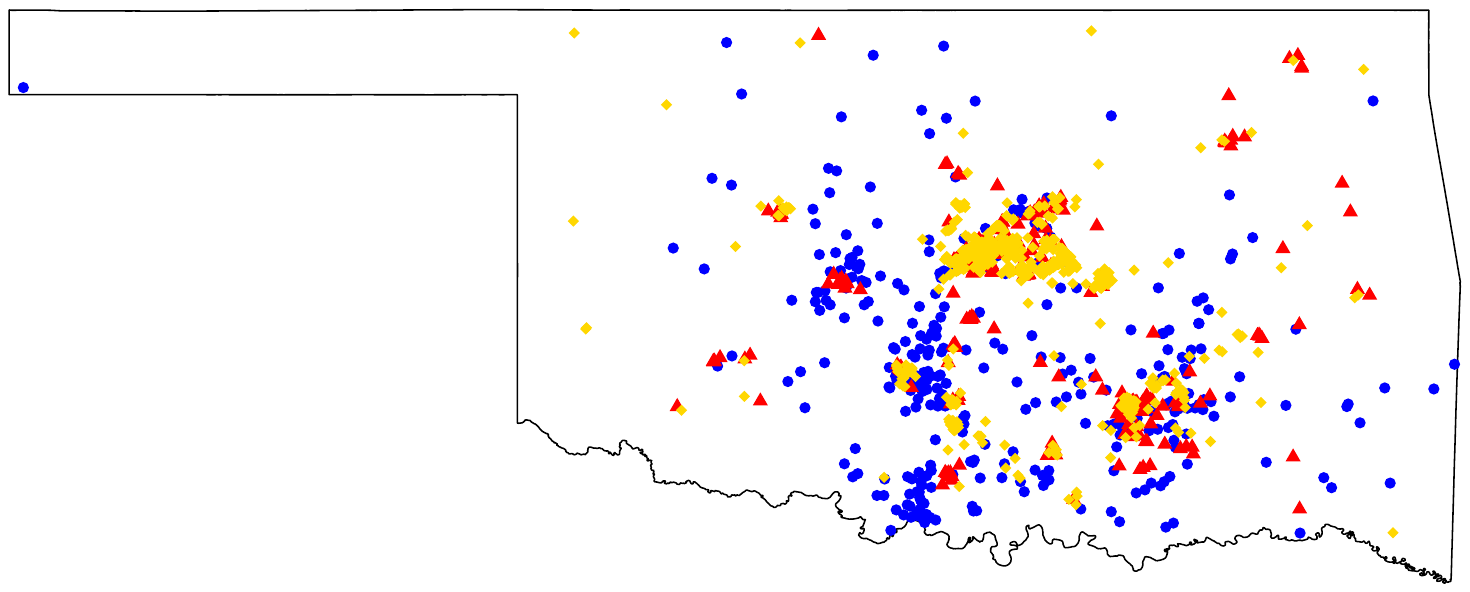}
\label{fig:earthquakes}}
\subfigure[]
{\includegraphics[width = 0.45\linewidth]{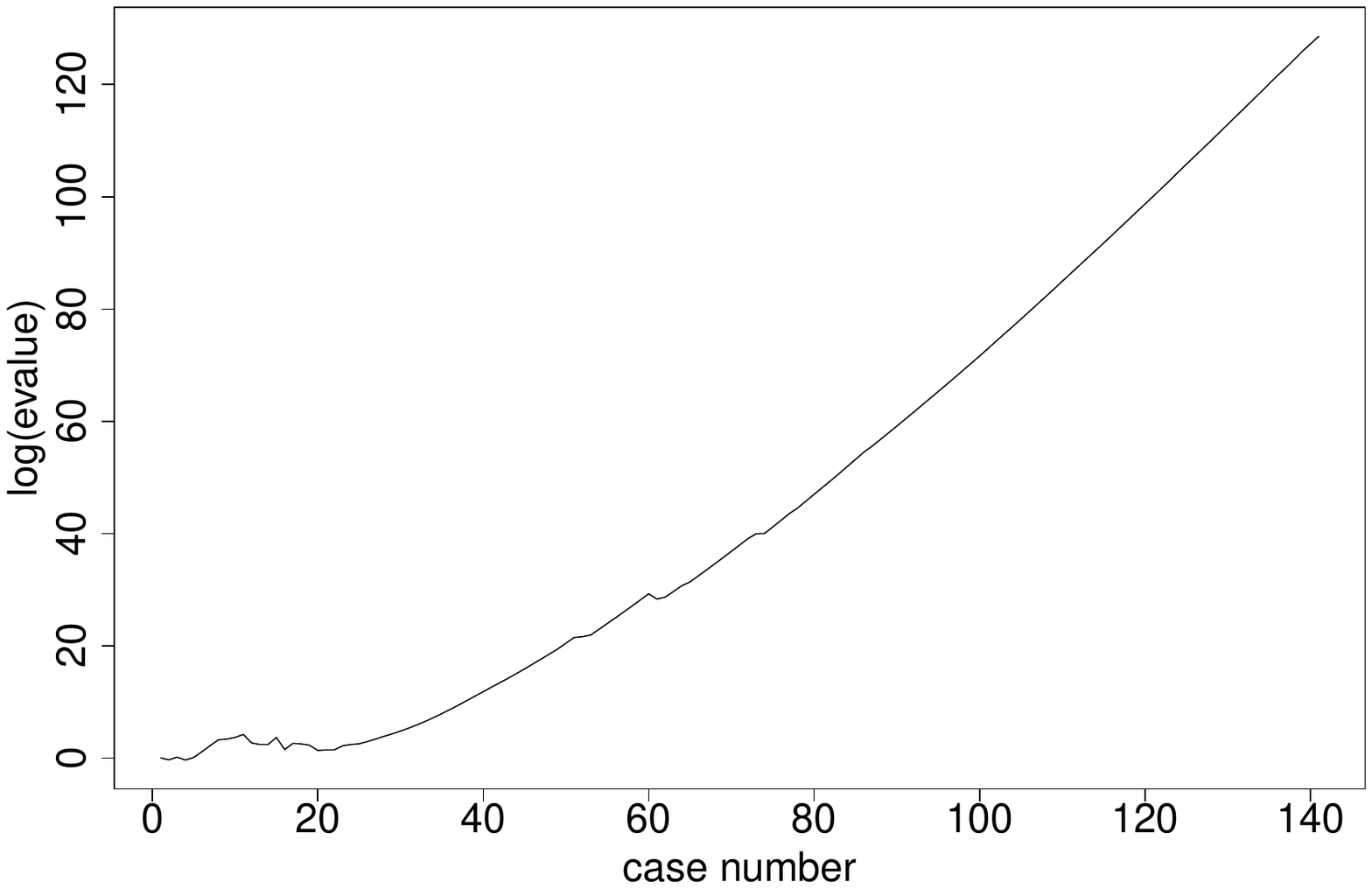}
\label{fig:earthquake_3_all}}
\subfigure[]
{\includegraphics[width = 0.45\linewidth]{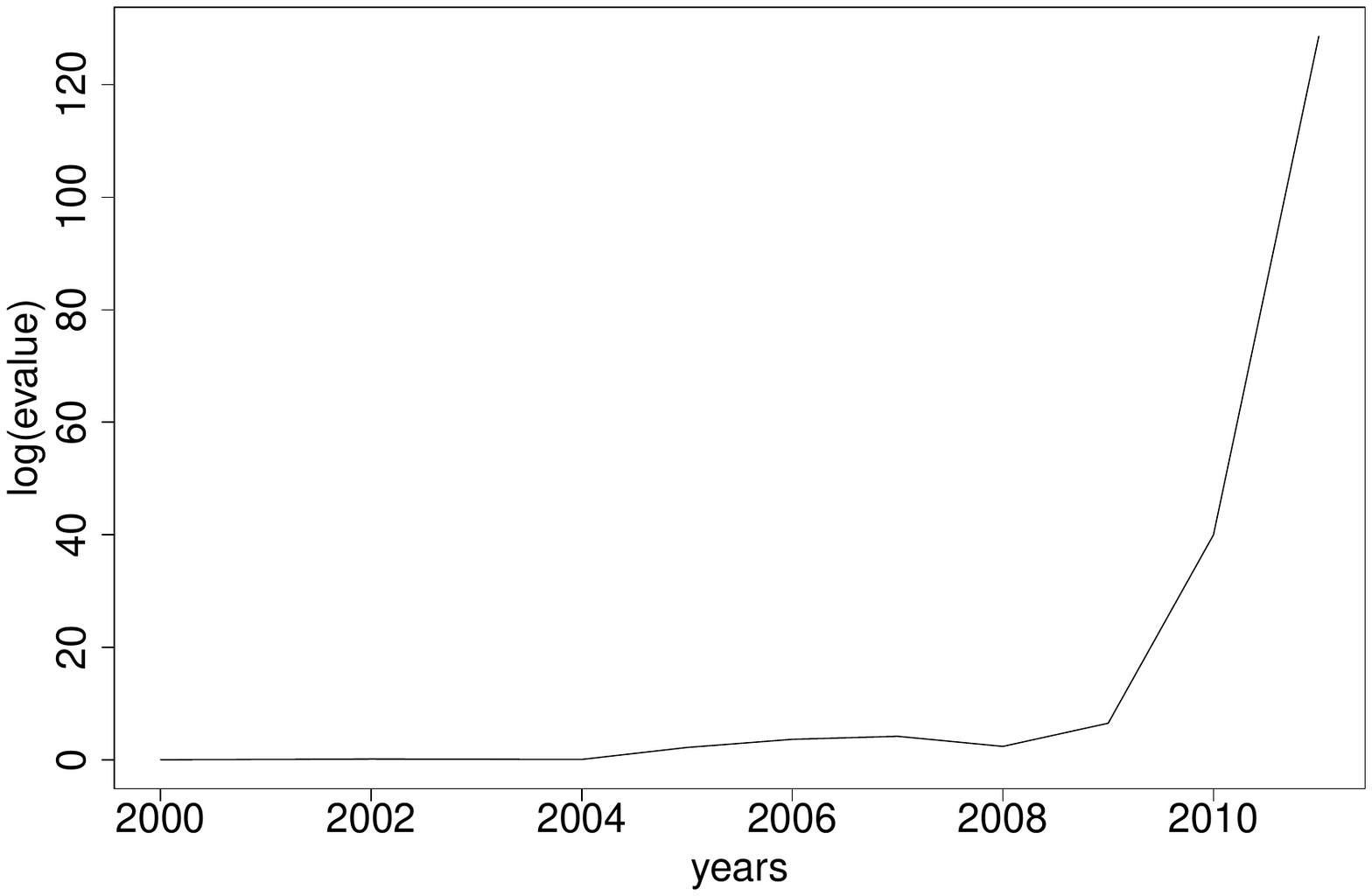}
\label{fig:earthquake_3_years}}
\caption{(a) Locations of all earthquakes (irrespective of magnitude) in Oklahoma from 2000 to 2011, with those in 2000 - 2009 (blue), 2010 (red) and 2011 (yellow), (b) Plot of $\log E_n^\text{\sc pr}$ vs index of occurrence (c) Plot of $\log E_n^\text{\sc pr}$ vs (year).}
\end{figure}

\end{illus}

\section{Discussion}

Methodology in anytime inference is a relatively new and evolving area in the statistical literature. In this paper, we introduce this concept to the spatial statistics domain, addressing the problem of testing for complete spatial randomness (CSR). While the problem of testing for CSR has been widely studied, our proposed approach is novel, particularly in its ability to handle online data streams. This is notable because traditional methods for testing CSR assume that all data points are available at once, whereas our approach is uniquely suited for real-time data, which is increasingly common in spatio-temporal applications such as environmental monitoring, epidemiology, and urban planning.

To construct our testing framework, we proposed the use of a beta mixture model to characterize the alternative hypothesis. As discussed in the main text, this choice is not the only possible formulation and more importantly not a limitation of the method. As the mixture density is modeled using the predictive recursion (PR) algorithm, which offers considerable flexibility, any well-behaved kernel can be incorporated into the mixture model. This adaptability allows for modifications to the PReprocess to accommodate different kernels, potentially improving the efficiency and robustness of the test. Our primary objective was to develop a general and practical online testing approach for CSR that can be applied to any point pattern. Through our experiments, we demonstrate that our method is both valid and efficient for diverse point processes like the Matérn process, a monotone inhomogeneous Poisson point process, and two real-world data applications.

Furthermore, our experiments hint at a natural connection between our sequential testing framework and change-point detection. Specifically, we suggest that our approach can be extended to e-detectors \citep{ramdasdetectors}, an online mechanism for detecting change-points, in spatial patterns. This extension is particularly relevant in the spatio-temporal context, where practitioners are not only interested in identifying clustering patterns but also in pinpointing the exact time or index points at which spatial structures undergo significant changes. Working on this aspect further would require additional theoretical developments, which we consider a promising direction for future research.

\section{Acknowledgements}
This article is released to inform interested parties of ongoing research and to encourage discussion. The views expressed on statistical issues are those of the authors and not those of the NSF or U.S. Census Bureau. This research was partially supported by the U.S. National Science Foundation (NSF) under NSF grants NCSE-2215168.

\appendix

\section{Algorithms}
\label{algo}

\begin{algorithm}[!]
\SetAlgoLined
\LinesNumbered
\SetKwInOut{Input}{input}\SetKwInOut{Output}{output}
\SetKwFunction{PRticlefilter}{PRticlefilter}
\Input {Region of interest $\Omega \subset \RR^2$, observations $s_1, \ldots, s_n$, initial guess $p_0$, random sample $U_1, \ldots, U_T$ from $p_0$, and weight sequence $\{w_i: i \geq 1\} \subset (0,1)$, kernel function $k(s \mid U_t)$} 
\Output {$\log {E_n^\text{\sc pr}}$} 
Rescale $\Omega$ to $\Omega^{*} \subset [0,1]^{2}$\;
Standardize observations $(s_1, \ldots, s_n)$ to $(s_1^{*}, \ldots, s_n^{*}) \in \Omega^{*}$\;
$\bm s_0$ $\leftarrow$ Generate a homogeneous point pattern with $\lambda = 20000$ on window = $\Omega^{*}$\;
$p^{\dagger} \leftarrow$ \PRticlefilter($s_0$, $k(s \mid U_t)$, $U_t$, $p_0$, $1$, $w$)\;
$(p_0^{\dagger} \,,\, D_0^{\dagger}) \gets \text{Extract from } p^{\dagger}$\;
Set $\log {E_n^\text{\sc pr}} =$ NULL and $L = 0$\;
 \For{$i=1,\ldots,n$}{
  pr $\leftarrow$ \PRticlefilter($s_i$, $k(s \mid U_t)$, $U_t$, $p_0^{\dagger}$, $D_0^{\dagger}, w$)\;
  $(p_0^{\dagger} \,, L^{\dagger} , \, D_0^{\dagger}) \gets$ \text{Extract from pr }\;
  $L \gets L - L^{\dagger}$\;
  $\text{Den}[i] \gets -i \log(\text{area of} \,\, \Omega^{*} )$\;
  $\log {E_n^\text{\sc pr}}[i] \gets L - \text{Den}[i]$\;
  }
 return $\log {E_n^\text{\sc pr}}$ .
 \caption{\textbf{PRe-process for testing CSR}}
 \label{algo:CSR}
\end{algorithm}

\begin{algorithm}[!]
\SetKwFunction{PRticlefilter}{PRticlefilter}
\DontPrintSemicolon
\SetAlgoLined
Initialize: Data $X_1, \ldots, X_n$, kernel function $k(x \mid U)$, initial guess $p_0$, random sample $U_1, \ldots, U_T$ from $p_0$, and weight sequence $\{w_i: i \geq 1\} \subset (0,1)$\;
Set $D_{0,t} = 1$ for $t=1,\ldots,T$ and $L=0$\;
\SetKwProg{Fn}{Function}{:}{}
\Fn{\PRticlefilter{$X, k, U, p_0, D_0, w$}}{
  \For{$i=1,\ldots,n$}{
  set $Nr_{t,i} = k(X_i \mid U_t) \, p_{i-1}(U_t)$ for each $t$, and $Dr_i = T^{-1}\sum_{t=1}^{T} k(X_i \mid U_t) \, D_{0,t}$\;
  update  $p_i(U_t) = (1 - w_i) p_{i-1}(U_t) + w_i \, Nr_{t,i} / Dr_i$ for each $t$\;
  evaluate $D_{0,t} = D_{0,t} [1 + w_i \{k(X_i \mid U_t) / Dr_i - 1\} ]$ for each $t$\;
  update $L = L - \log{Dr_i}$
  }
 return $p_n$, $D_{0}$, $L$.
}
\caption{\textbf{PRticle filter algorithm}}
\label{algo:prticle}
\end{algorithm}

\bibliographystyle{apalike}
\bibliography{refs}

\end{document}